\documentstyle[11pt]{article}
\normalbaselines

\begin{document}

\def\bea*{\begin{eqnarray*}}
\def\eea*{\end{eqnarray*}}
\def\ba{\begin{array}}
\def\ea{\end{array}}
% --------------------------------------------------------------
% If you want to see the names of the equations and references
% set below \count1=0 otherwise \count1=1
% --------------------------------------------------------------
\count1=1
% --------------------------------------------------------------
\def\be{\ifnum \count1=0 $$ \else \begin{equation}\fi}
\def\ee{\ifnum\count1=0 $$ \else \end{equation}\fi}
\def\ele(#1){\ifnum\count1=0 \eqno({\bf #1}) $$ \else \label{#1}\end{equation}\fi}
\def\req(#1){\ifnum\count1=0 {\bf #1}\else \ref{#1}\fi}
\def\bea(#1){\ifnum \count1=0   $$ \begin{array}{#1}
\else \begin{equation} \begin{array}{#1} \fi}
\def\eea{\ifnum \count1=0 \end{array} $$
\else  \end{array}\end{equation}\fi}
\def\elea(#1){\ifnum \count1=0 \end{array}\label{#1}\eqno({\bf #1}) $$
\else\end{array}\label{#1}\end{equation}\fi}
\def\cit(#1){
\ifnum\count1=0 {\bf #1} \cite{#1} \else 
\cite{#1}\fi}
\def\bibit(#1){\ifnum\count1=0 \bibitem{#1} [#1    ] \else \bibitem{#1}\fi}
\def\ds{\displaystyle}
\def\hb{\hfill\break}
\def\comment#1{\hb {***** {\em #1} *****}\hb }

\newcommand{\TZ}{\hbox{\bf T}}
\newcommand{\MZ}{\hbox{\bf M}}
\newcommand{\ZZ}{\hbox{\bf Z}}
\newcommand{\NZ}{\hbox{\bf N}}
\newcommand{\RZ}{\hbox{\bf R}}
\newcommand{\CZ}{\,\hbox{\bf C}}
\newcommand{\PZ}{\hbox{\bf P}}
\newcommand{\QZ}{\hbox{\rm eight}}
\newcommand{\HZ}{\hbox{\bf H}}
\newcommand{\EZ}{\hbox{\bf E}}
\newcommand{\GZ}{\,\hbox{\bf G}}

\font\germ=eufm10
\def\goth#1{\hbox{\germ #1}}
\vbox{\vspace{38mm}}

\begin{center}
{\LARGE \bf On ${\bf \tau^{(2)}}$-model in Chiral Potts Model and Cyclic Representation of  Quantum Group ${\bf U_q (sl_2)}$ } \\[10 mm] 
Shi-shyr Roan \\
{\it Institute of Mathematics \\
Academia Sinica \\  Taipei , Taiwan \\
(email: maroan@gate.sinica.edu.tw ) } \\[25mm]
\end{center}

\begin{abstract}
We identify the precise relationship between the five-parameter $\tau^{(2)}$-family in the $N$-state chiral Potts model and XXZ chains with $U_q (sl_2)$-cyclic representation. By studying the Yang-Baxter relation of the six-vertex model, we discover an  
one-parameter family of $L$-operators in terms of the quantum group $U_q (sl_2)$. 
When $N$ is odd, the $N$-state $\tau^{(2)}$-model can be regarded as the XXZ chain of $U_{\sf q} (sl_2)$ cyclic representations with ${\sf q}^N=1$. The symmetry algebra of the $\tau^{(2)}$-model is described by the quantum affine algebra $U_{\sf q} (\widehat{sl}_2)$ via the canonical representation. In general for an arbitrary $N$, we show that the XXZ chain with a $U_q (sl_2)$-cyclic representation for $q^{2N}=1$ is equivalent to two copies of the same $N$-state $\tau^{(2)}$-model. 
\end{abstract}
\par \vspace{5mm} \noindent
{\rm 2008 PACS}:  05.50.+q, 02.20.Uw, 75.10Jm \par \noindent

\setcounter{section}{0}
\section{Introduction}
\setcounter{equation}{0}
In the study of $N$-state chiral Potts model (CPM), Bazhanov and Stroganov \cite{BazS} discovered a five-parameter family of $L$-operators of $\tau^{(2)}$-model in the six-vertex model with a particular field (see \cite{B049} page 3), i.e. the solution of Yang-Baxter (YB) equation for the {\it asymmetric} six-vertex $R$-matrix, (see (\req(L)) and (\req(Rtau)) in this paper). 
The chiral Potts transfer matrix can be constructed as the Baxter's $Q$-operator of the $\tau^{(2)}$-model \cite{BBP, R0710}. By the functional relations between the fusion matrices of $\tau^{(2)}$-model and the chiral Potts transfer matrix, one can compute the eigenvalue spectrum of the (homogeneous) superintegrable CPM  \cite{AMP, B93, B94}, where the $\tau^{(2)}$-degeneracy occurs with the symmetry algebra described by Onsager algebra \cite{R05o}. Furthermore, one can discuss the eigenvalue spectrum and calculate the order parameter in CPM through functional-relation method \cite{B91, B90, B05, MR, R0805}. Though much progress has been made on issues related to the eigenvalues, we still lack enough information at present about the eigenvectors in CPM. Consequently, some important problems which require the knowledge of eigenvectors, such as the calculation of correlation functions, remain unsolved in the theory. On the other hand, the ("zero-field") six-vertex model (\cite{B04} Sect. 3) with the symmetric $R$-matrix has been a well-studied theory with profound knowledge known in literature about their eigenvalues and eigenvectors. The understanding of the structure was further extended to XXZ models of higher spin \cite{KiR}, or more generally the model associated to a representation of the quantum group $U_q(sl_2)$. When $q$ is a $N$th root of unity, it was shown in \cite{DFM, De05, FM01, NiD, R05b, R06F} that the degeneracy of XXZ models occurs with the extra $sl_2$-loop-algebra symmetry induced from the structure of the quantum affine algebra $U_q(\widehat{sl}_2)$. For odd $N$, as a special member of a one-parameter $U_q(sl_2)$-cyclic representation, the spin-$(N-1)/2$ XXZ chain can be identified with  the superintegrable $N$-state $\tau^{(2)}$-model \cite{R075}. Hence the superintegrable $\tau^{(2)}$-model also carries the $sl_2$-loop-algebra symmetry for the degeneracy (of certain sectors), compatible with the Onsager-algebra symmetry inherited from the chiral Potts transfer matrix \cite{R075}. The interrelation of these two symmetries can lead to some useful information about eigenvectors of the superintegrable model (see, e.g. \cite{AuP7, AuP8}). The aim of this paper is to extend the equivalent relationship between the $N$-state $\tau^{(2)}$-model and XXZ model with  $U_q(sl_2)$-cyclic representation to the whole five-parameter $\tau^{(2)}$-family found in \cite{BazS} for any $N$. Through the YB relation in the six-vertex model, we find a characterization of quantum group $U_q(sl_2)$ through a one-parameter $L$-operators of the YB solution (see (\req(6vL)), (\req(6YB)) in the paper). When $q$ is a root of unity, the algebra $U_q (sl_2)$  possesses a three-parameter family of cyclic representation. Together with the rescaling of the spectral parameter, XXZ chains with $U_q (sl_2)$-cyclic representation also carry a five-parameter $L$-operators.  By extending the argument in \cite{R075}, we show the identification of $\tau^{(2)}$-models and the XXZ chains with $U_q (sl_2)$-cyclic representation.  Note that such a precise connection between $\tau^{(2)}$-family in CPM and that XXZ-family with symmetric $R$-matrix seems not to have appeared in the literature before, to the best of the author's knowledge, even in odd $N$ case except a one-parameter family of  $\tau^{(2)}$-models in \cite{R075}. 
We hope an identification of these models will help to provide some useful clues to understand the eigenvectors of CPM in certain special cases.

This paper is organized as follows. In section \ref{sec:Uqsl}, we recall the definition of $\tau^{(2)}$-model,   quantum group $U_q (sl_2)$, and the quantum affine algebra $U_q (\widehat{sl}_2)$. In section \ref{sec:XXZ}, we first describe the three-parameter family of cyclic representations of $U_q (sl_2)$ for a root of unity $q$. The case of 
${\sf q}^N=1$ for odd $N$ is discussed in subsection \ref{ssec.oddN} where the $\tau^{(2)}$-model is identified with the XXZ chain with cyclic representation of $U_{\sf q} (sl_2)$, hence the quantum space carries a canonical representation of the quantum affine algebra $U_{\sf q} (\widehat{sl}_2)$.  In subsection \ref{ssec.allN}, we study the $N$-state $\tau^{(2)}$-model for an arbitrary $N$. We show that a XXZ chain of $U_q (sl_2)$-cyclic representation with $q^{2N}=1$ is equivalent to two copies of the same $N$-state $\tau^{(2)}$-model. Finally we close in section \ref{sec.F} with a brief concluding remark.

Notation: We use standard notations. For a positive integer $N$ greater than one, 
$\CZ^N$ denotes the vector space of $N$-cyclic vectors with the canonical base 
$|n \rangle, n \in \ZZ_N ~ (:= \ZZ/N\ZZ)$. We fix the $N$th root of unity $\omega = {\rm e}^{\frac{2 \pi {\rm i}}{N}}$, and  $X, Z$, the Weyl $\CZ^N$-operators :
$$
 X |n \rangle = | n +1 \rangle , ~ \ ~ Z |n \rangle = \omega^n |n \rangle ~ ~ \ ~ ~ (n \in \ZZ_N) ,
$$
which satisfy $X^N=Z^N=1$ and the Weyl relation: $XZ= \omega^{-1}ZX$.

\section{$N$-state $\tau^{(2)}$-model and  quantum group $U_q (sl_2)$  \label{sec:Uqsl}}
\setcounter{equation}{0}
The $N$-state $\tau^{(2)}$-model \cite{BBP, BazS, R0805} (also called the Baxter-Bazhanov-Stroganov model \cite{GIPS}) is the five-parameter family of $L$-operators of $\CZ^2$-auxiliary, $\CZ^N$-quantum space with entries expressed by Weyl operators $X, Z$:
\be
{\tt L} ( t ) = \left( \begin{array}{cc}
        1  -  t \frac{{\sf c}  }{\sf b' b} X   & (\frac{1}{\sf b }  -\omega   \frac{\sf a c }{\sf b' b} X) Z \\
       - t ( \frac{1}{\sf b'}  -  \frac{\sf a' c}{\sf b' b} X )Z^{-1} & - t \frac{1}{\sf b' b} + \omega   \frac{\sf a' a c }{\sf b' b} X
\end{array} \right) ,
\ele(L)
where $t$ is the spectral variable, and ${\sf a, b, a', b', c}$ are non-zero complex parameters. It is known that the $L$-operator (\req(L)) satisfies the YB equation 
$$
R(t/t') ({\sf L} (t) \bigotimes_{aux}1) ( 1
\bigotimes_{aux} {\tt L} (t')) = (1
\bigotimes_{aux} {\tt L} (t'))({\tt L} (t)
\bigotimes_{aux} 1) R(t/t') 
$$
for the asymmetry six-vertex $R$-matrix
\be
R(t) = \left( \begin{array}{cccc}
        t \omega - 1  & 0 & 0 & 0 \\
        0 &t-1 & \omega  - 1 &  0 \\ 
        0 & t(\omega  - 1) &( t-1)\omega & 0 \\
     0 & 0 &0 & t \omega - 1    
\end{array} \right).
\ele(Rtau)
The monodromy matrix, $
\bigotimes_{\ell=1}^L  {\tt L}_\ell (t)$ where ${\tt L}_\ell (t)= {\tt L}(t)$ at  site $\ell$, again satisfies the above YB relation, whose $\omega$-twisted trace defines the $\tau^{(2)}$-matrix: 
\be
\tau^{(2)}(t) = {\rm tr}_{\CZ^2} \bigotimes_{\ell=1}^L  {\tt L}_\ell (\omega t),
\ele(tau2) 
commuting with the spin-shift operator $X (:= \prod_{\ell} X_\ell)$.

The quantum group $U_q (sl_2)$ is the associated $\CZ$-algebra generated by $K^\frac{\pm 1}{2}, e^{\pm}$ with the relations $K^\frac{1}{2} K^\frac{-1}{2} = K^\frac{- 1}{2} K^\frac{1}{2} =1$  and
\be 
 K^{\frac{1}{2}} e^{\pm } K^\frac{-1}{2}  =   q^{\pm 1} e^{\pm}  , ~ 
~ [e^+ , e^- ] = \frac{K-K^{-1}}{q - q^{-1}} .
\ele(Uq)
Then the two-by-two matrix with $U_q (sl_2)$-entries defines a one-parameter family of $L$-operators:
\be
{\cal L} (s)   =  \left( \begin{array}{cc}
         \rho^{-1} s K^\frac{-1}{2}   -  s^{-1} K^\frac{1}{2}   &  (q- q^{-1}) e^-    \\
        (q - q^{-1}) e^+ &  s K^\frac{1}{2} -  \rho  s^{-1} K^\frac{-1}{2} 
\end{array} \right), ~ ~ \ \rho \neq 0 \in \CZ  ,
\ele(6vL)
which satisfy the YB equation 
\be
R_{\rm 6v} (s/s') ({\cal L}(s) \bigotimes_{aux}1) ( 1
\bigotimes_{aux} {\cal L}(s')) = (1
\bigotimes_{aux} {\cal L}(s'))( {\cal L}(s)
\bigotimes_{aux} 1) R_{\rm 6v} (s/s').
\ele(6YB)
for the six-vertex (symmetric) $R$-matrix \cite{Fad, KRS}:
$$
R_{\rm 6v} (s) = \left( \begin{array}{cccc}
        s^{-1} q - s q^{-1}  & 0 & 0 & 0 \\
        0 &s^{-1} - s & q - q^{-1} &  0 \\ 
        0 & q -q^{-1} &s^{-1} - s & 0 \\
     0 & 0 &0 & s^{-1} q - s q^{-1} 
\end{array} \right) .
$$   
Indeed, the YB relation (\req(6YB)) for ${\cal L}$ in (\req(6vL)) is the necessary and sufficient condition of the constraint (\req(Uq)) in the definition of quantum group $U_q (sl_2)$. Using the local $L$-operator (\req(6vL)), one constructs the monodromy matrix
 $\bigotimes_{\ell=1}^L {\cal L}_\ell (s)$ with entries in $( \stackrel{L}{\bigotimes} U_q (sl_2)) (s) $,
$$
\bigotimes_{\ell=1}^L {\cal L}_\ell (s)  =  \left( \begin{array}{cc}
        {\cal A}_L (s)  & {\cal B}_L (s) \\
        {\cal C}_L (s) & {\cal D}_L (s)
\end{array} \right) 
$$
again satisfying (\req(6YB)). The structure of the quantum affine algebra $U_q ( \widehat{sl}_2)$ is described by the leading and lowest terms of the above entries in  the monodromy matrix,  
$$
\begin{array}{ll}
{\sf A}_+ = \lim_{s  \rightarrow \infty} ( \rho^{-1} s)^{- L}{\cal A}_L(s) , & 
{\sf A}_- = \lim_{s \rightarrow 0} (- s)^L {\cal A}_L(s) ,   \\ 
{\sf B}_\pm = \lim_{s^{\pm 1} \rightarrow \infty} (\pm s)^{\mp (L-1)} \frac{{\cal B}_L(s)}{ q-q^{-1}} ,  & 
{\sf C}_\pm = \lim_{s^{\pm 1} \rightarrow \infty} (\pm s )^{\mp (L-1)}\frac{{\cal C}_L(s)}{q-q^{-1}},  \\
 {\sf D}_+ = \lim_{s  \rightarrow \infty}  s^{- L}{\cal D}_L(s), & {\sf D}_- = \lim_{s \rightarrow 0} (- \rho^{-1} s)^L {\cal D}_L(s). 
\end{array}  
$$ 
One has ${\sf A}_+ = {\sf A}_-^{-1}  , {\sf A}_\mp = {\sf D}_\pm  $. Denote $T^- = {\sf B}_+,  S^- = {\sf B}_-$ and $ S^+ = {\sf C}_+, T^+ = {\sf C}_-$.  The operators 
$$
k_0^{-1} =k_1  = {\sf A}^2_- = {\sf D}^2_+  , ~ \ e_1  = S^+ , \  f_1  = S^-,  ~ \ e_0 = T^-  , ~  f_0 = T^+  
$$
generate the quantum affine algebra $U_q(\widehat{sl}_2)$ with the Hopf-algebra structure
$$
\begin{array}{ll}
\bigtriangleup (k_i) = k_i \otimes k_i, & i=0, 1 , \\
\bigtriangleup (e_1 ) =  k_1 \otimes e_1    +   \rho^{- 1} e_1 \otimes k_0 , &
\bigtriangleup (f_1 ) =  k_1 \otimes f_1    +  \rho f_1 \otimes k_0 , \\
\bigtriangleup (e_0 ) =  \rho^{-1} k_0 \otimes e_0    +  e_0 \otimes k_1 , &
\bigtriangleup (f_0 ) =  \rho k_0 \otimes f_0    +  f_0 \otimes k_1 , \\
\end{array}
$$
Indeed, the explicit expression of generators of $U_q(\widehat{sl}_2)$ is 
\bea(l)
{\sf A}_- = K^{\frac{1}{2}} \otimes \cdots \otimes K^{\frac{1}{2}}, \ ~ ~ \ ~
{\sf A}_+ = K^{ \frac{-1}{2}} \otimes \cdots \otimes K^{\frac{-1}{2}}, \\
S^\pm = \sum_{i=1}^L  \underbrace{K^{\frac{1}{2}} \otimes \cdots \otimes K^{\frac{1}{2}}}_{i-1}\otimes e^\pm \otimes  \underbrace{K^{ \frac{-1}{2}} \otimes \cdots \otimes K^{ \frac{-1}{2}}}_{L-i} \rho^{\mp (L-i)} , 
\\
T^\pm  =  \sum_{i=1}^L  \rho^{\pm (i-1)} \underbrace{K^{\frac{-1}{2}} \otimes \cdots \otimes K^{ \frac{-1}{2}}}_{i-1}\otimes e^\pm \otimes  \underbrace{K^{\frac{1}{2}} \otimes \cdots \otimes K^{\frac{1}{2}}}_{L-i} . 
\elea(STpm)
Note that (\req(6YB)) is still valid when changing the variable $s$ by $\alpha s$ for a non-zero complex $\alpha$. Given a finite-dimensional representation $\sigma : U_q (sl_2) \longrightarrow {\rm End} (\CZ^d)$, the $L$-operator of $\CZ^2$-auxiliary and $\CZ^d$-quantum space, $L(s) = \sigma ({\cal L} (\alpha s))$, satisfies the YB relation (\req(6YB)). In particular, by setting the parameter $\rho =1 $ in (\req(6vL)), with $\alpha = q^{\frac{d-2}{2}}$ and $\sigma$ the spin-$\frac{d-1}{2}$ (highest-weight) representation of $U_q (sl_2)$ on  $\CZ^d = \oplus_{k=0}^{d-1} \CZ {\bf e}^k $:
$$
K^{\frac{1}{2}} ({\bf e}^k) = q^{\frac{d-1-2k}{2}} {\bf e}^k , \ \ e^+ ( {\bf e}^k ) = [ k  ] {\bf e}^{k-1} , \ \ e^-( {\bf e}^k ) = [ d-1-k ] {\bf e}^{k+1},
$$
where $[n]= \frac{q^n - q^{-n}}{q- q^{-1}}$ and $ e^+ ( {\bf e}^{0} ) = e^- ( {\bf e}^{d-1} )= 0$, one obtains the well-known $L$-operator of XXZ chain of spin-$\frac{d-1}{2}$ (see, e.g. \cite{KiR, R06Q, R06F} and references therein).

\section{The equivalence of $\tau^{(2)}$-models and XXZ chains with cyclic representation of  $U_q (sl_2)$ \label{sec:XXZ}}
\setcounter{equation}{0}
In this section, we consider the case when $q$ is a root of unity. For a $N$th root of unity $q$, there exists a three-parameter family of cyclic representation $\sigma_{\phi, \phi^\prime, \varepsilon}$ of $U_q ( sl_2)$, labeled by non-zero complex numbers $\phi, \phi^\prime$ and $\varepsilon$, which acts on cyclic $\CZ^N$-vectors  by
\bea(ll)
K^\frac{1}{2} |n \rangle =  q^{-n+\frac{\phi^\prime-\phi}{2}} |n \rangle , & \\
e^+  |n \rangle = q^\varepsilon \frac{  q^{\phi + n}-  q^{- \phi  - n}  }{q-q^{-1}} |n - 1 \rangle, &
e^-  |n \rangle = q^{-\varepsilon} \frac{ q^{\phi^\prime  - n}-  q^{- \phi^\prime + n}  }{q-q^{-1}} |n + 1 \rangle ,
\elea(crep)
(see, e.g. \cite{DJMM, DK}). 
The $L$-operator, 
\be
L(s) = \sigma_{\phi, \phi^\prime, \varepsilon} {\cal L}(s)
\ele(XXZL) 
gives rise to the transfer matrix of the XXZ chain with the $U_q (sl_2)$-cyclic representation $\sigma_{\phi, \phi^\prime,  \varepsilon}$,
\be
T (s) =   ( \otimes \sigma_{\phi, \phi^\prime,  \varepsilon} ) ( {\rm tr}_{\CZ^2} \bigotimes_{\ell=1}^L {\cal L}_\ell  (s))  ,
\ele(TcXZ)
which commutes with $K^\frac{1}{2} (:= \otimes_\ell K^\frac{1}{2}_\ell$, the product of local $K^\frac{1}{2}$-operators ).  

\subsection{The identification of the $N$-state $\tau^{(2)}$-model and the XXZ chain of $U_{\sf {\sf q}} (sl_2)$-cyclic representation  with ${\sf q}^N=1$ for odd $N$   \label{ssec.oddN}}
In the subsection, we consider the case $N$ odd, and write $N= 2M+1$.  Let ${\sf q}$ be the primitive $N$th root-of-unity with ${\sf q}^{-2} = \omega$. Then one can express the cyclic representation (\req(crep)) in terms of the Weyl operators $X, Z$: 
$$
\begin{array}{lll}
K^\frac{1}{2} = {\sf q}^\frac{\phi^\prime- \phi}{2}  Z^\frac{1}{2} , & 
e^+  = {\sf q}^{\varepsilon} \frac{(  {\sf q}^{\phi+1} Z^{- \frac{ 1}{2}} -  {\sf q}^{- \phi -1} Z^{  \frac{ 1}{2}})X^{- 1}}{{\sf q}-{\sf q}^{-1}} ,  &
e^-  = {\sf q}^{-\varepsilon} \frac{(  {\sf q}^{\phi^\prime +1} Z^{  \frac{ 1}{2}} -  {\sf q}^{- \phi^\prime -1} Z^{- \frac{ 1}{2}})X}{{\sf q}-{\sf q}^{-1}} ,
\end{array}
$$
hence follow the expression:
\bea(ll)
K^{-1} = {\sf q}^{\phi - \phi^\prime}  Z^{-1} , & \\
K^\frac{-1}{2} e^+  = -{\sf q}^{\frac{-\phi - \phi^\prime}{2}+\varepsilon-1} \frac{( 1 -{\sf q}^{2\phi +2}Z^{-1}   )X^{- 1}}{{\sf q}-{\sf q}^{-1}},  &
K^\frac{-1}{2}e^-  = {\sf q}^{\frac{\phi+\phi^\prime}{2}-\varepsilon+1} \frac{(  1   -  {\sf q}^{- 2\phi^\prime -2}Z^{-1} )X}{{\sf q}-{\sf q}^{-1}}.
\elea(eXZ)
Using the Fourier basis $\widehat{| n \rangle} = \frac{1}{N} \sum_{j \in \ZZ_N} \omega^{-nj} |j \rangle$, one may convert the Weyl operator $(Z^{-1}, X)$ to $(X, Z)$:  
\be
\bigg( {}^{Z^{-1}}_X \widehat{|0 \rangle}, \cdots, {}^{Z^{-1}}_X \widehat{| N-1 \rangle} \bigg) = \bigg( \widehat{|0 \rangle}, \cdots, \widehat{| N-1 \rangle} \bigg) {}^X_{Z} , 
\ele(Ft)
hence represent $K^{-1}, K^\frac{-1}{2} e^\pm$ in (\req(eXZ))   by  
$$
\begin{array}{lll}
K^{-1} = {\sf q}^{\phi - \phi^\prime} X , & 
K^\frac{-1}{2} e^+  =- {\sf q}^{\frac{-\phi - \phi^\prime}{2}+ \varepsilon-1 } \frac{( 1 -{\sf q}^{2\phi +2}X   )Z^{- 1}}{{\sf q}-{\sf q}^{-1}}, &
K^\frac{-1}{2} e^-  = {\sf q}^{\frac{\phi+\phi^\prime}{2} - \varepsilon+1} \frac{(  1  -  {\sf q}^{- 2\phi^\prime -2}X )Z}{{\sf q}-{\sf q}^{-1}}. 
\end{array}
$$ 
By the above expression and introducing the spectral variable $t = \lambda^{-1} s^2$ for a non-zero complex $\lambda$, the modified $L$-operator of (\req(XXZL)), $-s K^{\frac{-1}{2}}L (s)$, is gauge equivalent to
$$
\left( \begin{array}{cc}
        1- t \lambda \rho^{-1}  {\sf q}^{\phi - \phi^\prime}      X   & {\sf q}^{-\varepsilon} ( 1  -  {\sf q}^{- 2 \phi^\prime -2} X ) Z   \\
        - t   \lambda  {\sf q}^{\varepsilon} (1 -   {\sf q}^{2 \phi + 2} X ) Z^{- 1} & - t \lambda   +   \rho {\sf q}^{\phi - \phi^\prime}    X
\end{array} \right) ,
$$
which is the same as the $L$-operator (\req(L)) of $\tau^{(2)}$-model by the following identification of parameters:
\bea(l)
{\sf a }  = \lambda^{-1} \rho   {\sf q}^{- \phi - \phi^\prime -\varepsilon}  , ~ {\sf a'} = \rho  {\sf q}^{\phi + \phi^\prime + \varepsilon+2} , ~ {\sf b } = {\sf q}^{\varepsilon}, ~  {\sf b'}  = \lambda^{-1} {\sf q}^{-\varepsilon}, ~ {\sf c}  = \rho^{-1} {\sf q}^{\phi - \phi^\prime} ,  \\ 
{\rm equivalently} ~ ~ ~ ~ {\sf q}^{\varepsilon} = {\sf b } , ~ {\sf q}^{2\phi}= \frac{\omega {\sf a' c}}{\sf b} , ~ {\sf q}^{2 \phi'}= \frac{\sf b'}{\sf a c} , ~ \rho^2 = \frac{\omega {\sf a a'}}{\sf b b'}, ~ \lambda=  \frac{1}{\sf b b'} . 
\elea(par)
This implies the transfer matrix (\req(TcXZ)) of XXZ chain is equivalent to the $\tau^{(2)}$-transfer matrix (\req(tau2)).
Note that the product of local operator $K^{-1}$'s of XXZ chain is now corresponding to a scalar multiple of the spin-shift operator $X$, which commutes with the $\tau^{(2)}$-matrix. The representations  with $\varepsilon=0, \phi = \phi^\prime$ in (\req(crep)) form the one-parameter cyclic representation of $U_{\sf q} ( sl_2)$ discussed in \cite{R075} section 4. In particular, the case $\phi = \phi^\prime = M$ is the spin-$\frac{N-1}{2}$ highest-weight representation of $U_{\sf q} (sl_2)$.

It is known that the degeneracy of $\tau^{(2)}$-model in CPM occurs in the alternating superintegrable case (see, \cite{R0805} section 4.3), where the $\tau^{(2)}$-model is characterized by the $L$-operator (\req(L)) with the parameter 
$$
{\sf a } = \omega^{\rm m} {\sf b'}, ~ {\sf a'} = \omega^{\rm m'} {\sf b }, ~ {\sf c}= \omega^{\rm n} , ~ ~ ({\rm m}, {\rm m'}, {\rm n} \in \ZZ),
$$
or equivalently,  with the representation parameters $\phi, \phi', \varepsilon$ and $\rho, \lambda$ in (\req(par)) given by 
\be
\phi   = -({\rm m' +n }+1) , ~ \phi^\prime = {\rm m} + {\rm  n }, ~ {\sf q}^{\varepsilon}= {\sf b} , ~ \rho = {\sf q}^{-(1+{\rm m}+{\rm m'})}, ~  \lambda= \frac{1}{\sf b b'}.
\ele(alsupar)
In this case, there exist the normalized $N$th power of $S^\pm , T^\pm$ in (\req(STpm)), $S^{\pm (N)}= \frac{S^{\pm N}}{[N]!},  T^{\pm (N)}= \frac{S^{\pm N}}{[N]!}$. As $\phi $ in (\req(alsupar)) is an integer, by the algebraic-Bethe-ansatz technique, one can show the degeneracy of the alternating superintegrable $\tau^{(2)}$-model (for certain sectors) possesses the symmetry algebra generated by $S^{\pm (N)}, T^{\pm (N)}$, which can be identified with the $sl_2$-loop algebra under certain constraints between the integers ${\rm m}, {\rm m}'$ and ${\rm n}$, as in the discussion of the homogeneous superintegrable case in \cite{DFM, NiD, R06F}. However, the precise structure about the symmetry algebra for arbitrary ${\rm m}, {\rm m}', {\rm n}$ remains unknown to be identified.

\subsection{XXZ chain of $U_q (sl_2)$-cyclic representation  with $q^{2N}=1$ as two copies of  the $N$-state $\tau^{(2)}$-model for an arbitrary $N$  \label{ssec.allN}}
By extending the argument of the previous subsection, we now identify the $N$-state $\tau^{(2)}$-model with a XXZ chain for an arbitrary $N$. Let $q$ be a primitive $(2N)$-th root of unity with $q^{-2} = \omega$, hence $q^N=-1$. Consider the three parameter family of $U_q (sl_2)$-cyclic representations $\sigma_{\phi, \phi^\prime, \varepsilon}$ in (\req(crep)) on the vector space $V$ of $2N$-cyclic vectors:
$$
V : = \bigoplus \{  \CZ |n \rangle' | n \in \ZZ_{2N} \} .
$$ 
Denote the vectors in $V$: 
$$
\begin{array}{l}
|n \rangle : = |n \rangle' + |n +N \rangle' ,  ~ ~ (n \in \ZZ_{2N}) , \\
|n \rangle_- : = |n \rangle' - |n +N \rangle' , ~ ~ ( 0 \leq n \leq N-1). 
\end{array}
$$
Then $|n \rangle  = |n+N \rangle $. We define $|n \rangle_-$ for $n \in \ZZ$ by the $N$-periodic condition:  $|n +N \rangle_- = |n \rangle_-$. Hence we can identify $\{ |n \rangle \}$ or $\{ |n \rangle_-\}$ with the canonical basis of the vector space of $N$-cyclic vectors, and one has the decomposition of $V$:
$$
V = V_+ \oplus V_- , ~ ~ V_+ := \sum_{n \in \ZZ_N} \CZ |n \rangle , \ ~ ~  V_- := \sum_{n \in \ZZ_N} \CZ |n \rangle_- .
$$
The spin-shift operator $|n \rangle' \mapsto |n+1 \rangle'$ of $V$ is decomposed as the sum of $X$ on $V_+$ and $\widetilde{X}$ on $V_-$, where 
$$
X|n \rangle = |n+1 \rangle ~ \ (n \in \ZZ_N ) , ~~~~
\widetilde{X}|n \rangle_- = \left\{ \begin{array}{ll}
   |n+1 \rangle_- & n \not \equiv N-1  \pmod{N},   \\
   -|n+1 \rangle_- & n \equiv N-1 \pmod{N}.\\
\end{array} \right.
$$
Together with the operator $Z$: $Z|n \rangle = \omega^n |n \rangle$, $Z|n \rangle_- = \omega^n |n \rangle_-$, one finds $(X, Z)$ , $(\widetilde{X}, Z )$ are Weyl operators of $V_+$ or $V_-$ respectively with $X^N = Z^N = 1, \widetilde{X}^N = -1$.
Under the cyclic representation $\sigma_{\phi, \phi^\prime, \varepsilon}$ (\req(crep)) of $U_q(sl_2)$, the subspaces $V_+, V_-$ are interchanged under the generators $K^\frac{1}{2}, e^\pm$, with the expression: 
\bea(lll) 
K^\frac{1}{2} | n \rangle  = q^{-n+\frac{\phi^\prime- \phi }{2}} | n \rangle_- , & 
K^\frac{1}{2} | n \rangle_-  = q^{-n+\frac{\phi^\prime- \phi }{2}} | n \rangle  & ( 0 \leq n \leq N-1 ) , \\
e^+  | n \rangle = q^{\varepsilon } \frac{  q^{\phi + n}-  q^{-\phi - n}  }{q-q^{-1}} | n - 1 \rangle_-  & (1 \leq n \leq N),  \\
e^- | n \rangle = q^{-\varepsilon }  \frac{q^{\phi^\prime - n}- q^{-\phi^\prime + n}  }{q-q^{-1}}| n +1 \rangle_-     & (-1 \leq n  \leq N-2) , 
\\

e^+ | n \rangle_- = q^{\varepsilon } \frac{  q^{\phi + n}-  q^{-\phi - n}  }{q-q^{-1}} | n-1 \rangle , &
e^- | n \rangle_-= q^{-\varepsilon }\frac{q^{\phi^\prime - n}- q^{-\phi^\prime + n}  }{q-q^{-1}}| n +1 \rangle &( 0 \leq n  \leq N-1) .
\elea(QVpm)
Using the above formulas, one finds that $K^{-1}$ and $K^\frac{-1}{2}e^\pm $ are operators of  $V_+$ and $V_-$, which can be expressed by $X, Z$, and $\widetilde{X}, Z$ respectively such that the relation (\req(eXZ)) (using $q$) holds for $V_+$, and also valid for $V_-$ when replacing $X$ by $\widetilde{X}$.  By using the $V_-$-basis, $| n \rangle \rangle : = q^n | n \rangle_-$ for $ 0 \leq n \leq N-1$, the  Weyl pair  $(\widetilde{X}, Z)$ is converted to  $(q^{-1} X, Z)$. One can express the operators $K^\frac{-1}{2}e^\pm $ by 
$$
\begin{array}{ll}
K^\frac{-1}{2}  e^+ | n \rangle \rangle = - q^{\frac{-\phi - \phi^\prime}{2}+\varepsilon} \frac{(1 -q^{2\phi +2}Z^{-1}   )X^{- 1}}{q-q^{-1}}| n \rangle \rangle ,
&
K^\frac{-1}{2} e^- | n \rangle \rangle = q^{\frac{\phi+\phi^\prime}{2} -\varepsilon} \frac{( 1   - q^{- 2\phi^\prime -2}Z^{-1} )X}{q-q^{-1}} | n \rangle \rangle. 
\end{array}
$$
Using the Fourier transform (\req(Ft)) of the basis $\{  |n \rangle \}$ and $\{ |n \rangle \rangle \}$, by the same argument in subsection \ref{ssec.oddN}, one can show that 
$-s K^{\frac{-1}{2}}L (s)$ for the $L$-operator (\req(XXZL)) is gauge equivalent to two copies of the same $L$-operator (\req(L)) of $\tau^{(2)}$-model with the parameter given by (\req(par)) (replacing ${\sf q}$ by $q$).  Therefore the transfer matrix (\req(TcXZ)) of XXZ chain is equivalent to the sum of two copies of $\tau^{(2)}$-transfer matrix (\req(tau2)). Note that  neither one of the two copies of $\tau^{(2)}$-model inherits the $U_q(sl_2)$ structure of  XXZ chain, in which one copy is sent to another by (\req(QVpm)). The symmetry structure of $\tau^{(2)}$-model related to $U_q(sl_2)$ is different from that in subsection \ref{ssec.oddN}  for the odd $N$ case.  Nevertheless, the $\tau^{(2)}$-model with certain constraints on the parameter again possesses the $sl_2$-loop-algebra symmetry that arises from the $U_q(\widehat{sl}_2)$-structure of the corresponding XXZ-model. In particular, in the homogeneous superintegrable case, the Onsager-algebra symmetry of the degenerate $\tau^{(2)}$-eigenspace is extended to the symmetry of $sl_2$-loop algebra (in certain sectors). The relationship of these two symmetries will hopefully, though not immediately apparent, lead to a solution of the eigenvector problem in the superintegrable CPM  (for an arbitrary $N$) along the line in  \cite{AuP7, AuP8}.

\section{Concluding Remarks}\label{sec.F} 
By studying the general solution of YB equation of the six-vertex model, we find a one-parameter $L$-operators associated to the quantum group $U_q (sl_2)$, which carries a three-parameter family of cyclic representation when $q$ is a root of unity. We have showed that the XXZ chain of  $U_q (sl_2)$-cyclic representation  with $q^{2N}=1$ is equivalent to the sum of two copies of a $N$-state $\tau^{(2)}$-model. In particular, the Onsager-algebra symmetry of the homogeneous superintegrable CPM is enlarged to the $sl_2$-loop-algebra symmetry induced from the corresponding XXZ-model. When $N$ is odd and ${\sf q}$ is a primitive $N$th root of unity, the  $N$-state $\tau^{(2)}$-model can also be identified with the XXZ chain of a
cyclic representation of $U_{\sf q} (sl_2)$, hence the quantum space carries the structure of quantum affine algebra $U_{\sf q} (\widehat{sl}_2)$. As a special case, the homogeneous superintegrable $\tau^{(2)}$-model in CPM is equivalent to the spin-$(N-1)/2$ XXZ chain with the canonical $U_{\sf q} (\widehat{sl}_2)$-structure \cite{R075}, by which the $sl_2$-loop-algebra symmetry of the superintegrable $\tau^{(2)}$-model was derived in \cite{NiD, R06F}. In the recent study of  eigenvalues of the chiral Potts model with alternating rapidities \cite{R0805}, the $\tau^{(2)}$-degeneracy is also found in the alternating superintegrable case. By identifying these $\tau^{(2)}$-models with the XXZ chains via (\req(alsupar)), we find the $sl_2$-loop-algebra symmetry for the degenerate $\tau^{(2)}$-eigenspace in certain cases.  
Similar to the homogeneous CPM case \cite{R05o}, some twisted version of Onsager-algebra symmetry for the $\tau^{(2)}$-model could also possibly exist in the theory of a general alternating superintegrable chiral Potts model. 
However, the explicit nature remains hard to know precisely. The exact relationship and a further symmetry study of general cases are now under consideration.

\section*{Acknowledgements}
The author wishes to thank Professor S. Kobayashi for the invitation of visiting U. C. Berkeley in the spring of 2008,  where part of this work was carried out.
This work is supported in part by National Science Council of Taiwan under Grant No NSC  96-2115-M-001-004.


\begin{thebibliography}{99}
\bibitem{AMP} G. Albertini, B. M. McCoy, and 
J. H. H. Perk, Eigenvalue spectrum of the
superintegrable chiral Potts model, In {\it Integrable system in quantum field theory and statistical mechanics,}  Adv. Stud. Pure Math., 19, Kinokuniya Academic, Academic Press, Boston, MA (1989) 1--55.
%
\bibitem{AuP7} H. Au-Yang and J.H.H. Perk, Eigenvectors the superintegrable model I: $\goth{sl}_2$ generators, J. Phys. A: Math. Theor. 41 (2008) 275201; arXiv: 0710.5257.
%
\bibitem{AuP8} H. Au-Yang and J.H.H. Perk, Eigenvectors in the superintegrable model II: ground state sector, arXiv: 0803.3029,
%
\bibitem{B91} R. J. Baxter, Calculation of the eigenvalues of the transfer matrix of the chiral Potts model, {\it Proc. Fourth Asia-Pacific Physics Conference} (Seoul, Korea, 1990) Vol 1, World-Scientific, Singapore (1991) 42--58.
%
\bibitem{B90} R. J. Baxter, Chiral Potts model: eigenvalues of the transfer matrix, Phys. Lett. A 146 (1990) 110--114.
% 
\bibitem{B93} R. J. Baxter, Chiral Potts model with skewed boundary conditions, J.
Stat. Phys. 73 (1993) 461--495.
% 
\bibitem{B94} R. J. Baxter, Interfacial tension of the chiral Potts model, J. Phys. A: Math. Gen. 27 (1994) 1837--1849.
%
\bibitem{B04} R. J. Baxter, The six and eight-vertex models revisited, J. Stat. Phys. 114 (2004) 43--66; cond-mat/0403138.
%
\bibitem{B049} R. J. Baxter, Transfer matrix functional relation for the generalized $\tau_2(t_q)$ model, J. Stat. Phys. 117 (2004) 1--25;  cond-mat/0409493.
%
\bibitem{B05} R. J. Baxter, Derivation of the order parameter of the chiral Potts model, Phys.Rev.Lett. 94 (2005) 130602; cond-mat/0501227.
%
\bibitem{BBP} R. J. Baxter, V.V. Bazhanov and J.H.H. Perk,  Functional relations for transfer
matrices of the chiral Potts model, Int. J. Mod. Phys. B 4 (1990) 803--870.
%
\bibitem{BazS} V.V. Bazhanov and Yu.G. Stroganov, Chiral
Potts model as a descendant of the six-vertex model, J.
Stat. Phys. 59 (1990) 799--817.
%
\bibitem{DJMM} E. Date, M. Jimbo, K. Miki and T. Miwa, Cyclic representations of $U_q(sl(n+1, \CZ))$ at $q^N=1$, Publ. RIMS, Kyoto Univ. 27 (1991) 347--366.
%
\bibitem{DK} C. DeConcini and V. G. Kac, Representations of quantum groups at roots of unity, in {\it Operator Algebra, Unitary Representations, Enveloping Algebras, and Invariant Theory}, Paris (1989) {\it Progress in Mathematics} 92, Birkh\"{a}user, Boston, Massachusstts (1990) 471-- 506.
%
\bibitem{DFM} T. Deguchi, K. Fabricius and B. M. McCoy, The $sl_2$ loop algebra symmetry for the six-vertex model at roots of unity, J. Stat. Phys. 102 (2001) 701--736; cond-mat/9912141. 
%
\bibitem{De05} T. Deguchi: Regular XXZ Bethe states at roots of unity- as highest weight vectors of the $sl_2$ loop algebra at roots of unity, cond-mat/0503564 v3. 
%
\bibitem{FM01} K. Fabricius and B. M. McCoy, Evaluation parameters and Bethe roots for the six vertex model at roots of unity, {\it Progress in Mathematical Physics} Vol 23, eds. M. Kashiwara and T. Miwa,  Birkh\"{a}user Boston (2002), 119--144; cond-mat/0108057.
%
\bibitem{Fad} L. D. Faddeev, How algebraic Bethe Ansatz works for integrable models, eds. A.
Connes, K. Gawedzki and J. Zinn-Justin, {\it Quantum symmetries/ Symmetries quantiques},
Proceedings of the Les Houches summer school, Session LXIV, Les Houches, France, August 1-September 8, 1995, North-Holland, Amsterdam (1998),  149--219.
%
\bibitem{GIPS} G. von Gehlen, N. Iorgov, S. Pakuliak and V. Shadura: Baxter-Bazhanov- Stroganov model: Separation of variables and Baxter equation, J. Phys. A: Math. Gen. 39 (2006) 7257--7282; nlin.SI/0603028.
%
\bibitem{KiR} A, N. Kirillov and N. Yu. Reshetikhin, Exact solution of the integrable XXZ Heisenberg model with arbitrary spin: I. The ground state and the excitation spectrum, J. Phys. A: Math. Gen. 20 (1987) 1565 -- 1595.
%
\bibitem{KRS} P. P. Kulish,  N. Yu. Reshetikhin and E. K. Sklyanin, Yang Baxter equation and representation theory, Lett. Math. Phys. 5 (1981) 393--403.
%
\bibitem{MR} B. M. McCoy and S. S. Roan, Excitation spectrum and phase structure of the chiral Potts model. Phys. Lett. A 150 (1990) 347--354.
%
\bibitem{NiD} A. Nishino and T. Deguchi, The $L(sl_2)$ symmetry of the Bazhanov-Stroganov model associated with the superintegrable chiral Potts model, Phys. Lett. A 356 (2006) 366--370
; cond-mat/0605551.
%
\bibitem{R05o} S. S. Roan, The Onsager algebra symmetry of $\tau^{(j)}$-matrices in the superintegrable chiral Potts model, J. Stat. Mech. (2005) P09007; cond-mat/0505698.
%
\bibitem{R05b} S. S. Roan, Bethe ansatz and symmetry in superintegrable chiral Potts model and root-of-unity six-vertex model, in Nankai Tracts in Mathematics Vol. 10, {\it Differential Geometry and Physics}, eds. Mo-Lin Go and Weiping Zhang,  World Scientific, Singapore (2006), 399-409; cond-mat/0511543.
%
\bibitem{R06Q} S. S. Roan, The Q-operator for root-of-unity symmetry in six vertex model, J. Phys. A: Math. Gen. 39 (2006) 12303-12325; cond-mat/0602375.
%
\bibitem{R06F} S. S. Roan, Fusion operators in the generalized $\tau^{(2)}$-model and root-of-unity symmetry of the XXZ spin chain of higher spin, J. Phys. A: Math. Theor. 40 (2007) 1481-1511; cond-mat/0607258.
%
\bibitem{R075} S. S. Roan, The transfer matrix of superintegrable chiral Potts model as the Q-operator of root-of-unity XXZ chain with cyclic representation of $U_q(sl_2)$, J. Stat. Mech. (2007) P09021; arXiv: 0705.2856.
%
\bibitem{R0710} S. S. Roan, On the equivalent theory of the generalized $\tau^{(2)}$-model and the chiral Potts model with two alternating vertical rapidities, arXiv: 0710.2764.
%
\bibitem{R0805} S. S. Roan, Bethe equation of $\tau^{(2)}$-model and eigenvalues of finite-size transfer matrix of chiral Potts model with alternating rapidities, J. Stat. Mech. (2008) P10001; arXiv:0805.1585.
\end{thebibliography}
\end{document}